\documentclass[twoside,twocolumn,9pt]{article}
\usepackage{extsizes}
\usepackage[super,sort&compress,comma]{natbib} 
\usepackage[version=3]{mhchem}
\usepackage[left=1.5cm, right=1.5cm, top=1.785cm, bottom=2.0cm]{geometry}
\usepackage{balance}
\usepackage{mathptmx}
\usepackage{sectsty}
\usepackage{graphicx} 
\usepackage{lastpage}
\usepackage[format=plain,justification=justified,singlelinecheck=false,font={stretch=1.125,small,sf},labelfont=bf,labelsep=space]{caption}
\usepackage{float}
\usepackage{fancyhdr}
\usepackage{fnpos}
\usepackage[english]{babel}
\addto{\captionsenglish}{
}
\usepackage{array}
\usepackage{droidsans}
\usepackage{charter}
\usepackage[T1]{fontenc}
\usepackage[usenames,dvipsnames]{xcolor}
\usepackage{setspace}
\usepackage[compact]{titlesec}
\usepackage{hyperref}
\usepackage{subcaption}

\usepackage{epstopdf}

\definecolor{cream}{RGB}{222,217,201}

\begin{document}

\pagestyle{fancy}
\thispagestyle{plain}
\fancypagestyle{plain}{
\renewcommand{\headrulewidth}{0pt}
}

\makeFNbottom
\makeatletter
\renewcommand\LARGE{\@setfontsize\LARGE{15pt}{17}}
\renewcommand\Large{\@setfontsize\Large{12pt}{14}}
\renewcommand\large{\@setfontsize\large{10pt}{12}}
\renewcommand\footnotesize{\@setfontsize\footnotesize{7pt}{10}}
\makeatother

\renewcommand{\thefootnote}{\fnsymbol{footnote}}
\renewcommand\footnoterule{\vspace*{1pt}%
\color{cream}\hrule width 3.5in height 0.4pt \color{black}\vspace*{5pt}} 
\setcounter{secnumdepth}{5}

\makeatletter 
\renewcommand\@biblabel[1]{#1}            
\renewcommand\@makefntext[1]%
{\noindent\makebox[0pt][r]{\@thefnmark\,}#1}
\makeatother 
\renewcommand{\figurename}{\small{Fig.}~}
\sectionfont{\sffamily\Large}
\subsectionfont{\normalsize}
\subsubsectionfont{\bf}
\setstretch{1.125} 
\setlength{\skip\footins}{0.8cm}
\setlength{\footnotesep}{0.25cm}
\setlength{\jot}{10pt}
\titlespacing*{\section}{0pt}{4pt}{4pt}
\titlespacing*{\subsection}{0pt}{15pt}{1pt}


\fancyfoot{}
\fancyfoot[RO]{\footnotesize{\sffamily{1--\pageref{LastPage} ~\textbar  \hspace{2pt}\thepage}}}
\fancyfoot[LE]{\footnotesize{\sffamily{\thepage~\textbar\hspace{3.45cm} 1--\pageref{LastPage}}}}
\fancyhead{}
\renewcommand{\headrulewidth}{0pt} 
\renewcommand{\footrulewidth}{0pt}
\setlength{\arrayrulewidth}{1pt}
\setlength{\columnsep}{6.5mm}
\setlength\bibsep{1pt}

\makeatletter 
\newlength{\figrulesep} 
\setlength{\figrulesep}{0.5\textfloatsep} 

\newcommand{\topfigrule}{\vspace*{-1pt}%
\noindent{\color{cream}\rule[-\figrulesep]{\columnwidth}{1.5pt}} }

\newcommand{\botfigrule}{\vspace*{-2pt}%
\noindent{\color{cream}\rule[\figrulesep]{\columnwidth}{1.5pt}} }

\newcommand{\dblfigrule}{\vspace*{-1pt}%
\noindent{\color{cream}\rule[-\figrulesep]{\textwidth}{1.5pt}} }

\makeatother


\title{Role of softness on transition temperatures for pNIPAM microgels$^\dag$}

\author{Syamjith KS, Shubhasmita Rout and Alan R Jacob$^{\ast}$ }
\maketitle

\begin {abstract}
Poly(N-isopropylacrylamide) (pNIPAM) microgels are renowned for their thermoresponsive behavior, exhibiting a distinct volume phase transition (VPT) upon temperature changes. This study investigates the influence of microgel softness, controlled by varying the crosslinking density during synthesis via free radical polymerization (FRP), on the difference between the volume phase transition temperature (VPTT) and the electrokinetic transition temperature (ETT). These transition temperatures mark the points at which the microgel size and surface charge, respectively, undergo significant alterations in response to temperature. Here, we investigate this phenomenon, employing dynamic light scattering (DLS) and electrophoretic light scattering (ELS) measurements to characterize the size and electrophoretic mobility response of pNIPAM microgels with different crosslinking densities as a function of temperature. By analyzing the observed trends in the difference between the transition temperatures, we aim to develop a hypothesis that provides a deeper physical understanding of the microgel structure and its relationship to transition temperatures. This investigation thus sheds light on the intricate interplay between microgel structure and its thermoresponsive behavior, offering insights for the design and optimization of pNIPAM microgels for future applications.
\end{abstract}


\renewcommand*\rmdefault{bch}\normalfont\upshape
\rmfamily
\section*{}
\vspace{-1cm}


\footnotetext{\textit{$^{\ast}$~Soft Matter Group, Department of Chemical Engineering, Indian Institute of Technology Hyderabad, Kandi, Sangareddy, Telangana, India. 502284 \\ Tel: +91 4023016220; E-mail:arjacob@che.iith.ac.in}}

\footnotetext{\dag~Electronic Supplementary Information (ESI) available: [details of any supplementary information available should be included here]. See DOI: 10.1039/cXsm00000x/}



\section{Introduction}

Poly(N-isopropylacrylamide) (pNIPAM) is a polymer renowned for its thermoresponsive character. pNIPAM exhibits a Lower Critical Solution Temperature (LCST), a critical temperature above which polymer undergoes phase separation and precipitates out of solution \cite{deike2001rheology, crassous2006thermosensitive, senff1999rheology, heskins1968solution, kim2017nature, jain2015tunable, bischofberger2015new}. This observed temperature-dependent behavior can be attributed to changes in the energetic landscape surrounding the hydrophobic isopropyl groups within the pNIPAM polymer\cite{pelton2010poly}. At lower temperatures, the pNIPAM forms hydrogen bonds with water but as the temperatures increases, the energy required to maintain these hydrogen bonds between the pNIPAM and water molecules becomes untenable. Thus, to minimize this energetic penalty, the pNIPAM undergoes a conformational change wherein the polymer chains collapse and rearrange themselves in a way that minimizes their contact with surrounding water molecules. It's important to note that pNIPAM doesn't strictly become hydrophobic at higher temperatures; instead, it prioritizes minimizing unfavorable interactions with water, leading to the deswelling of pNIPAM. pNIPAM microgels \cite{saunders1999microgel,nayak2005soft, hoare2008characterizing} also demonstrate similar thermoresponsive behavior at a distinct volume phase transition temperature (VPTT) \cite{daly2000temperature, sennato2021double, del2021two, saunders2009microgels, pelton2000temperature, truzzolillo2018overcharging, vintha2020phase,saunders2004structure}. This refers to the specific temperature at which the microgel undergoes a shrinking due to the collapse of its polymer network. Microgels \cite{baker1949microgel} are classified as soft colloids (10$\mu m$>size>100$nm$), with three-dimensional polymer networks exhibiting unique properties due to their size and deformability. pNIPAM microgels, distinguished from smaller nanogels by their size (<100$nm$) \cite{karg2019nanogels,brijitta2009phase}, exhibit complex behaviors due to their pronounced network structure. These behaviors include volume phase transitions, morphological changes, and aggregation in response to environmental factors. Vlassopoulos and Cloitre \cite{vlassopoulos2014tunable}  in their review reported that soft colloids can be deformed under applied stress without rupturing \cite{stokes2008rheology, chen2010rheology, bonnecaze2010micromechanics, vlassopoulos2010polymers}.  Extensive research has explored the phase transition behaviour of pNIPAM microgels, highlighting their potential for various applications in waste water treatment, drug delivery, and other fields \cite{islam2014poly, hauck2022pnipaam, guan2011pNIPAM, sanzari2020poly, franco2021thermal}. The mechanical properties of pNIPAM microgel suspensions are significantly influenced by particle stiffness, surface properties, and inter particle interactions, as evidenced by their nonlinear viscoelastic behavior \cite{karthickeyan2016identification, zhu2024cation, vialetto2024effect, jijo2010volume,saisavadas2023large, karg2019nanogels,brijitta2009phase,fievet2001determining}.

Poly(N-isopropylacrylamide) (pNIPAM) microgels have been synthesized using two main methods: surfactant-free emulsion polymerization (SFEP), documented as early as 1986 \cite{saunders2009microgels, pelton1986preparation}, and free radical polymerization (FRP) with surfactants \cite{wu1994kinetics, braibanti2016impact, truzzolillo2018overcharging, still2013synthesis}. Both methods typically employ a reaction mixture containing the monomer N-isopropylacrylamide (NIPAM), an initiator, a crosslinker, and high-purity deionized water. It has been established that pNIPAM microgels possess a core-corona structure \cite{daly2000temperature,sengottiyan2023core,rivas2022link}. The core is a densely crosslinked inner region, while the corona is a more loosely crosslinked outer shell. Notably, the corona often contains charged functional groups that significantly influence the microgel's surface charge and its interactions with the surrounding environment \cite{del2021two}. The degree of crosslinking within the microgel can be controlled during synthesis by adjusting the crosslinker concentration. Higher crosslinker concentrations lead to a more rigid structure, effectively tuning the microgel's "softness." This ability to manipulate the crosslinking density plays a crucial role in microgel behavior.  Alternative synthesis techniques can achieve microgels with more homogeneous cross-link density distribution throughout the particle. For instance, a continuous feeding method demonstrated the feasibility of synthesizing mono disperse poly(N-isopropylacrylamide) microgel particles with uniform cross-linking \cite{acciaro2011investigation}. Theoretical studies employing the Debye-Hückel potential have shed light on how electrostatic interactions influence the overall conformation of microgels\textbf{ \cite{kobayashi2014structure}.}

The collapse of pNIPAM microgels upon heating is a complex process influenced by multiple factors. Recent studies have proposed a two-step deswelling mechanism \cite{del2021two} driven primarily by electrostatic interactions between charged groups located on the microgel corona. This model suggests an initial collapse of the core , followed by a subsequent contraction of the outer corona layer. While the three-step model proposed by  Daly and Saunders  \cite{daly2000temperature}, offers a more detailed description involving distinct transition states, the underlying principle of a sequential collapse of the microgel remains consistent. Both models underscore the importance of electrostatic effects and core-corona interactions in determining the overall deswelling behavior. Daly and Saunders \cite{daly2000temperature} designated the temperature at which the initial size reduction (core collapse) occurs as the VPTT, while the temperature associated with the increase in electrophoretic mobility (corona collapse) is termed the electrokinetic transition temperature (ETT). This increase in electrophoretic mobility is attributed to a rise in charge density due to the compaction of the corona during collapse. Daly and Saunders \cite{daly2000temperature} also observed that the VPTT was always 5-6°C lower than ETT.

The thermoresponsive nature of pNIPAM microgels have attracted a lot of attention from the research and industrial sectors. Their potential applications span diverse fields, including drug delivery \cite{hsiue2002preparation, choi2002galactosylated,murthy2002novel, nayak2004folate, nolan2005phase, lopez2005use, nolan2006h,hoare2008hydrogels}, bio-sensing \cite{islam2014poly, guan2011pNIPAM, retama2003microstructural}, microfluidics \cite{hauck2022pnipaam}, pharmaceuticals \cite{sanzari2020poly}, chemo-mechanical devices \cite{varga2006pulsating}, for the development of responsive interfaces \cite{acciaro2011investigation,serpe2003layer, schmidt2008packing, nerapusri2007absorption} and technologies for separation or purification  \cite{bromberg2003smart}. However, despite extensive research efforts, knowledge gaps persists regarding many aspects of pNIPAM microgels. This study aims to address one such gap by investigating the relationship between the difference between the ETT and VPTT vis a vis crosslinking density of the microgels or softness of the microgel. By systematically varying the crosslinking density during synthesis, we seek to elucidate how this key structural parameter influences the two distinct transition temperatures and the overall thermoresponsive behavior of pNIPAM microgels. Understanding this relationship will provide valuable insights for the design and optimization of pNIPAM-based materials with tailored thermoresponsive properties for specific applications.

The rest of this manuscript is arranged as follows. The "Materials and Methods" section details the materials used for pNIPAM microgel synthesis and the characterization techniques employed. This is followed by the "Results and Discussion" section, which has the data from experiments and analyzes of this data. Finally, the "Conclusion" section summarizes the key takeaways and potential implications of our research.

\section{Materials and methods}

\subsection{Chemicals} N-isopropylacrylamide (NIPAM,$C_6H_{11}NO,>98\%$), potassium persulfate ($K{_2}S{_2}O{_8},>97\%$), N,N'-methylenebisacrylamide (MBA,$C_7H_{10}N_2O_2,>99\%$) and sodium dodecyl sulfate (SDS,$NaC_{12}H_{25}SO_4,>98\%$) were procured from Sigma Aldrich and used as such without any modifications. Type 1 (milli-Q) water (resistivity of $18 M\Omega cm$) is used for all experiments.

\subsection{Microgel synthesis}
pNIPAM microgels with different crosslinker concentrations were synthesized by free-radical emulsion polymerization (FRP) with surfactant.  %

For  1$wt\%$ crosslinker concentration, 225$ml$ of type 1 milli-Q water was mixed with 2310$mg$ of NIPAM monomer, 24.4$mg$ of the crosslinker MBA and 24.2$mg$ of SDS to make monomer solution in a beaker. This monomer solution was agitated at 1000$rpm$ for 15$mins$.
A three-neck flask was submerged in a silicon oil bath (10$cSt$) placed on a hot plate stirrer heated to 85$^{\circ}C$ and set to 1000$rpm$. One of the neck is attached to a condenser connected to a chiller set at 10$^{\circ}C$ and another neck is connected for purging with inert gas. The monomer solution was transferred to the flask after the oil bath reached 75$^{\circ}C$. After the transfer of the reaction mixture, the reactor was purged with nitrogen for 30$mins$. A pre-prepared initiator solution, 90$mg$ of potassium persulfate (KPS) dissolved in 25$mL$ of type 1 water to synthesize anionic (negatively charged) pNIPAM microgels, was added to the three neck flask once the reaction mixture reaches 80$^{\circ}C$.

After addition of the initiator, the reaction mixture changed from translucent to turbid white after 10$mins$ which indicated the progress of the reaction. The reaction was carried out  for 4$hrs$ at 85$^{\circ}C$. Later the reaction mixture was allowed to cool overnight.

The reaction mixture was centrifuged a minimum of four cycles at 7400$rpm$ and 37$^{\circ}C$ for 90$mins$. After each cycle, the supernatant was removed and the sedimented microgel mixed with fresh type 1 water.
 
Additional batches were synthesized using the same methodology but with different crosslinker concentrations: 4$wt\%$, 5$wt\%$, 7.5$wt\%$, 10$wt\%$, and 12.5$wt\%$, respectively. Three replicates for each crosslinker concentration (total, 3*6 = 18 batches) were synthesized to ensure that the microgel characteristics were consistent.

\subsection{Dynamic light scattering (DLS) experiments}

Dynamic light scattering (DLS) is a non-destructive technique employed to investigate the hydrodynamic radius and temperature-dependent behavior of the synthesized pNIPAM microgels. The spherical morphology of the microgels was confirmed using scanning electron microscopy (SEM); the results are presented in the supplementary information (SI) \cite{si}. DLS measurements are particularly well-suited for characterizing the size distribution of spherical particles in suspension. An Anton Paar DLS (particle analyzer model LiteSizer 500), with a laser light source of 658$nm$ wavelength was used.

To assess the thermoresponsive behaviour of pNIPAM microgels, size measurements were performed at various temperatures during both heating and cooling cycles. 
The following experimental protocols were implemented:

1. Temperature profile: 
The temperature was changed in a step-wise manner, starting from 20$^{\circ}C$ and increasing to 50$^{\circ}C$ with increments of 2.5$^{\circ}C$. This process was then reversed for the cooling cycle, with measurements taken from 50$^{\circ}C$ back down to 20$^{\circ}C$ using the same 2.5$^{\circ}C$ step size. 

2. Equilibration Time: At each increment temperature the microgel sample was allowed to equilibrate for 10$mins$. This was done to ensure the sample reaches a steady state and the microgels responds to the set temperature.
 
The DLS data collected at each temperature point was analyzed using the cumulant approach by the software. This approach determines the average hydrodynamic diameter of the microgels in the sample, as well as the polydispersity index (PDI), which reflects the width of the size distribution. The cumulant analysis is appropriate for generally monodisperse spherical materials, such as the pNIPAM microgels used in this investigation. The DLS measurements were taken at a scattering angle of 90° (side scatter). This scattering angle is chosen because it provides excellent sensitivity for detecting these pNIPAM microgels.

\subsection{Electrophoretic light scattering (ELS) experiments}
ELS was used to investigate the electrophoretic mobility of the synthesized pNIPAM microgels. The ELS measurements were performed on the Anton Paar particle analyzer model LiteSizer 500 that had previously been employed for the DLS studies. 

An Omega cuvette integrated with gold electrodes was used for the ELS measurements. The presence of these electrodes facilitates the deployment of an electric field of 200 $V$ across the sample during the experiment.  This voltage generates an electric field within the sample, leading to the anionic pNIPAM microgels to migrate towards oppositely charged electrodes.

Electrophoretic mobility (ELS) measurements were performed using continuously-monitored phase-analysis light scattering (cmPALS). Unlike conventional ELS technologies, which rely primarily on light scattering and frequently require strong electric fields, cmPALS combines light scattering and real-time phase analysis. This method enables for mobility tests at lower electric fields, potentially eradicating Joule heating effects that may alter the characteristics of samples including pNIPAM microgels. Overall, cmPALS provides a more sensitive, reliable, and gentle assessment of electrophoretic mobility than conventional ELS methods.

\textbf{Sample preparation:~~}Both the DLS and ELS sample preparation are the same. 1$ml$ of stock solution of pNIPAM microgels was diluted with 14$ml$ type 1 water such that  the sample transmittance measured in the DLS was more than 30\%. This is done to reduce the possibility of multiple scattering events which have the potential to distort the  results.

\subsection{Data fits: Extracting critical temperatures}
The temperature sweep of DLS and ELS yielded information on the size (hydrodynamic diameter) and electrophoretic mobility of the pNIPAM microgels at different temperatures. Critical temperatures, such as the Volume Phase Transition Temperature (VPTT) and the Electrokinetic Transition Temperature (ETT), have to be extracted from the experimental data in order to more fully understand the microgel's thermoresponsive behavior. Three independent fitting methods were used  to elucidate a reliable determination of both volume phase and electrokinetic transition temperatures are described below.

\subsubsection{Half-way method:~~} This method leverages the assumption that the critical temperature (VPTT or ETT) coincides with the temperature at which the measured parameter (size or electrophoretic mobility) reaches the average value between its swollen and collapsed states of the microgels \cite{daly2000temperature}. 

\begin{equation}
\textrm{VPTT}  \ \textrm{or} \ \textrm{ETT} = T_a = T(y_a) 
\end{equation}
Where
$T_a$: Temperature at which the measured parameter ($y$ = size or electrophoretic mobility) is the average of its swollen ($y_s$) and collapsed ($y_b$) state values.

$y_a$: ($y_s + y_b$) / 2

This method offers a quick estimate of the transition temperatures but may have lower accuracy compared to the fitting methods.

\subsubsection{Modified Boltzmann sigmoidal model:~~} This method involves fitting the temperature-dependent data (e.g., size or electrophoretic mobility, both denoted by the function y) to a modified Boltzmann sigmoidal equation \cite{navarro2011modified, campos2023effective}. The equation used in this manuscript is as follows:
\begin{equation}
  y = y_b + ( \ y_b - y_s \ ) \ / \ ( \ W + exp \ (-\ (T - T_c) \ / \ \alpha))
\end{equation}

where
$y_b$: Plateau value of the function y in the collapsed state of microgel

$y_s$: Plateau value of the function y in the swollen state of microgel

$T_c$: Transition temperature (VPTT or ETT)

$W$: Characteristic width parameter of the transition

$\alpha$: The strength of transition (slope of the curve)

Non-linear regression algorithm fitting is  implemented in Orgin Pro software to determine the parameters that best match the experimental data. The VPTT or ETT can then be extracted from the fitted curve as the temperature corresponding to $T_c$.

\subsubsection{Phenomenological function analysis:~~} An alternative approach utilizes a phenomenological function specifically designed to describe the size or electrophoretic mobility changes of pNIPAM microgels response to temperature variations \cite{del2021two}. 

\begin{equation}
  y = y_0 - A\ tanh \ (s \ (T - T_c)+  A \ (T - T_c)
\end{equation}

where
$y_0$: y-intercept

$A$: Amplitude of the transition

$T_c$: Characteristic temperature of the transition (VPTT or ETT)

$s$: Characteristic width parameter of the transition

By fitting the experimental data to this function using software like Origin Pro, the VPTT or ETT can be obtained from the fitted parameter $T_c$.This method is advantageous if the specific form of the temperature dependence is known or can be reasonably assumed.

Employing all 3 analysis methods strengthens the determination of VPTT and ETT. The modified Boltzmann sigmoidal and phenomenological function, uses the mathematical function to fit a sigmoid curve to our experimental data. This s-shaped curve helps us identify the inflection point, which corresponds to the transition temperature. It is crucial to emphasize that all fitting parameters were allowed to freely vary during the optimization process, ensuring an unbiased determination of VPTT and ETT. The convergence of results from the fitting methods (modified Boltzmann sigmoidal and phenomenological function) with the half-way method provides increased confidence in the identifying the transition temperature. This approach helps minimize the influence of random errors or data fluctuations on the determination of this critical parameter. By applying these methods to the DLS and ELS data analyzed using Origin Pro, we expect to obtain reliable values for VPTT and ETT, which will be used to characterize the thermoresponsive behavior and stability of the synthesized pNIPAM microgels.

\section{Results and discussion}

\subsection{Reversible thermoresponsive nature}

\begin{figure}[h!]
\centering
  \includegraphics[height=6cm]{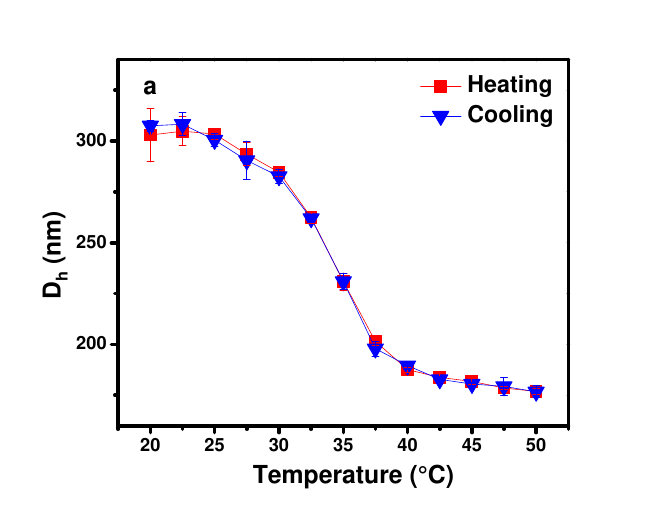}
   \includegraphics[height=6cm]{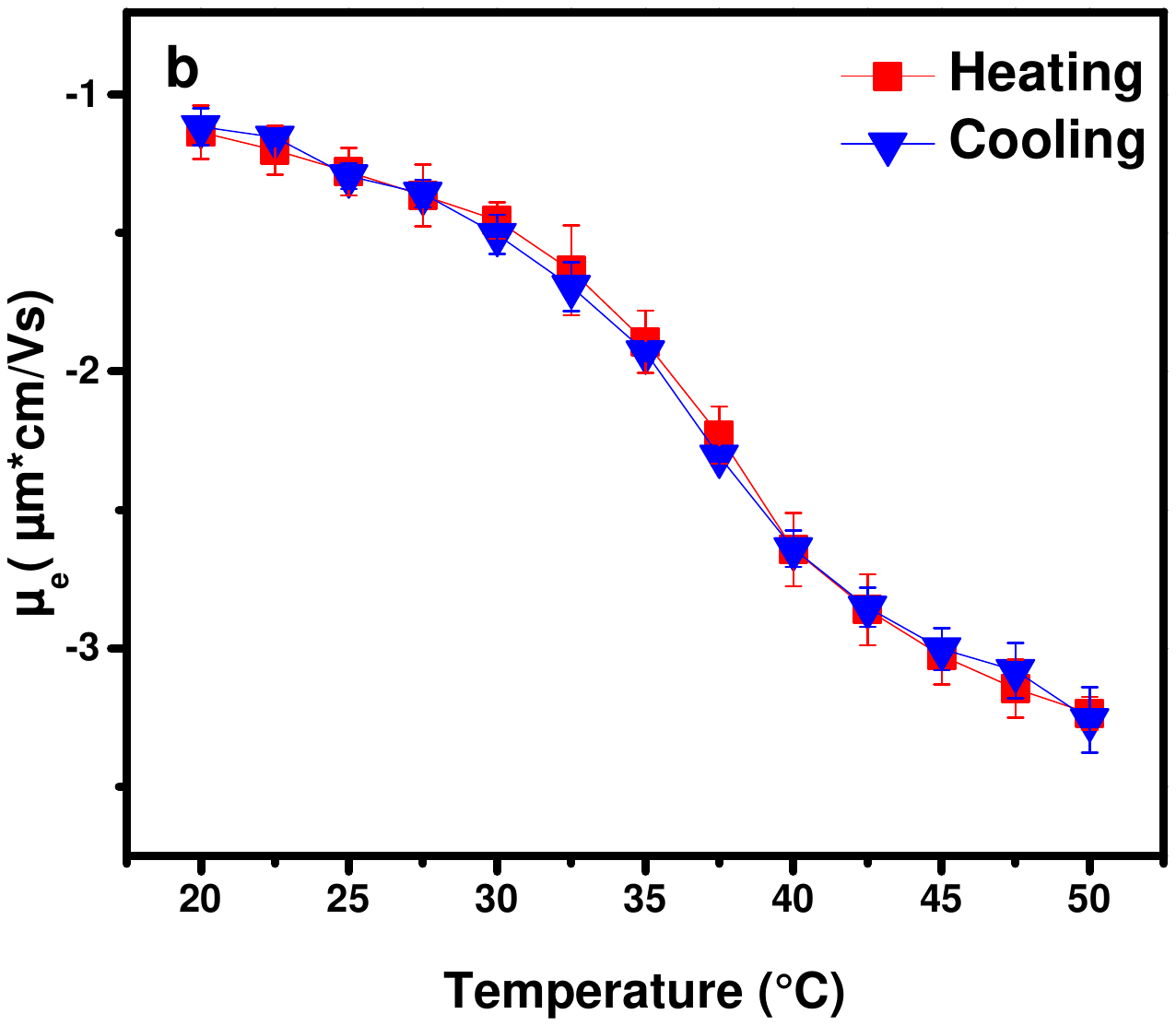}
  \caption{Reversible thermoresponsive behavior of pNIPAM microgels with 12.5$wt\%$ crosslinker: (a) hydrodynamic diameter and (b) electrophoretic mobility as a function of temperature. The heating and cooling cycles are represented by red (filled squares) and blue (inverted filled triangles), respectively.}
  \label{thermoreversibility}
\end{figure}

DLS and ELS temperature sweep experiments confirmed the reversible thermoresponsive behavior of all synthesized pNIPAM microgel batches \textit{(1$wt\%$, 4$wt\%$, 5$wt\%$, 7.5$wt\%$, 10$wt\%$, and 12.5$wt\%$ crosslinker concentrations)}. As shown by Fig. \ref{thermoreversibility} for the 12.5$wt\%$ crosslinker batch (and similar trends observed for all the other crosslinker concentrations presented in the supplementary information \cite{si}, both size (hydrodynamic diameter) and electrophoretic mobility (surface charge) exhibit a clear reversible trend during heating and cooling cycles. This aligns with well-established literature on pNIPAM microgels \cite{daly2000temperature, sennato2021double}

This volume phase transition leads to the reduction of hydrodynamic diameter with increasing temperature, as seen in Fig. \ref{thermoreversibility} (a). On the other hand, Fig. \ref{thermoreversibility} (b) reveals how the electrophoretic mobility of the microgels, which is related to the electrokinetic phase transition, albeit the surface charge density increases with increasing temperature. This increase in surface charge density is a consequence of the collapse of the polymer network in the microgel. On cooling, the pNIPAM chains re-hydrate and the microgels return to their original swollen state, with both size and mobility measurements recovering their initial values. This reversible response to temperature variations demonstrates the suitability of pNIPAM microgels for applications requiring stimuli-responsive behavior.

\subsection{Difference in volume phase and electrokinetic transition temperatures}

\begin{figure}[h!]
 \centering
 \includegraphics[height=6cm]{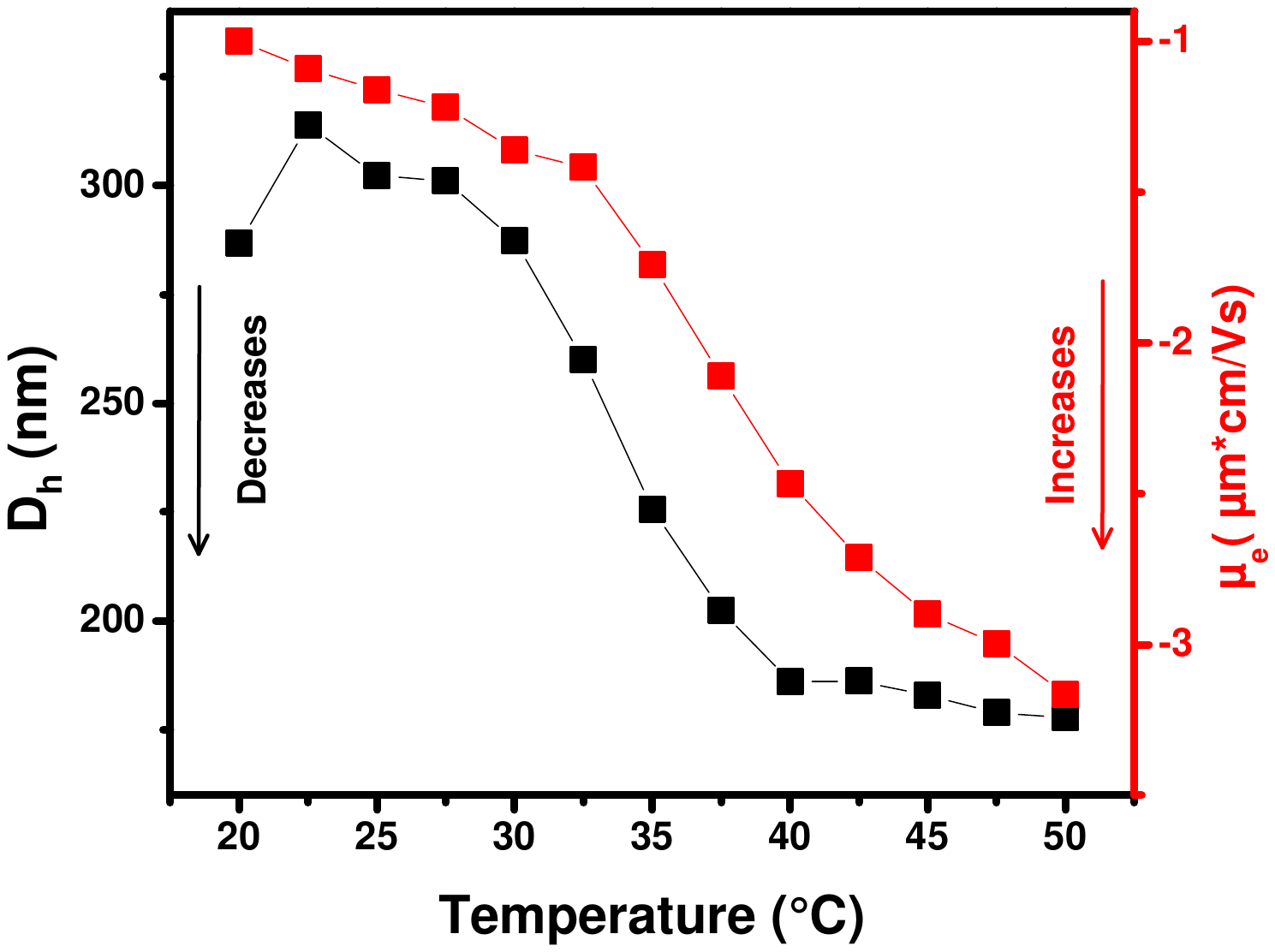}
 \caption{Distinguishing transitions: hydrodynamic diameter and electrophoretic mobility reveal difference in volume phase transition (VPTT) and electrokinetic transition temperature (ETT) for pNIPAM Microgels with 12.5 $wt\%$ crosslinker
: hydrodynamic diameter (left y-axis, black filled squares) and electrophoretic mobility (right y-axis, red filled squares) as a function of temperature. A clear difference is observed between the two transition temperatures.}
 \label{vpttekk}
\end{figure}

The double y-axis graph superimposes the hydrodynamic diameter and the electrophoretic mobility in Fig. \ref{vpttekk} with temperature. The hydrodynamic diameter and electrophoretic mobility decreases and increases respectively with the electrokinetic transition lagging in temperature compared to volume phase transition as seen in Fig. \ref{vpttekk}. It is interesting to note that during the cooling cycle follows the same pathway as the heating cycle from Fig. \ref{thermoreversibility}. This temperature lag in electrokinetic transition has been observed before and a theory was developed describing the individual structure of microgel as a core corona structure, \cite{daly2000temperature,del2021two} .

The core is a dense network of crosslinked pNIPAM chains and the corona a sparse polymer network comprising of ionic charges also. As the temperature increases, the core region which has the highest crosslinking density undergoes a collapse due to the conformational change of the polymer network. This initial collapse disrupts the hydration layer surrounding the core and initiates dehydration. However, the corona layer, with its lower crosslinking density with the ionic charges have a higher affinity for water molecules, exhibits a temperature delay in the collapse. Eventually, at a higher temperature, the corona layer also undergoes dehydration, leading to a complete shrinkage of the microgel and thus increasing the charge density of microgel. This process is known as the two-step collapse (heterogeneous collapse) \cite{del2021two} and explains the observed difference between the two transition temperatures i.e. VPTT and ETT. It is significant to remember that the microgel still contains less than 50\% water, even when it is collapsed \cite{varga2020effect}. 

\subsection{Effect of crosslinker concentration on size and mobility}

\begin{figure}[h!]
\centering
  \includegraphics[height=6cm]{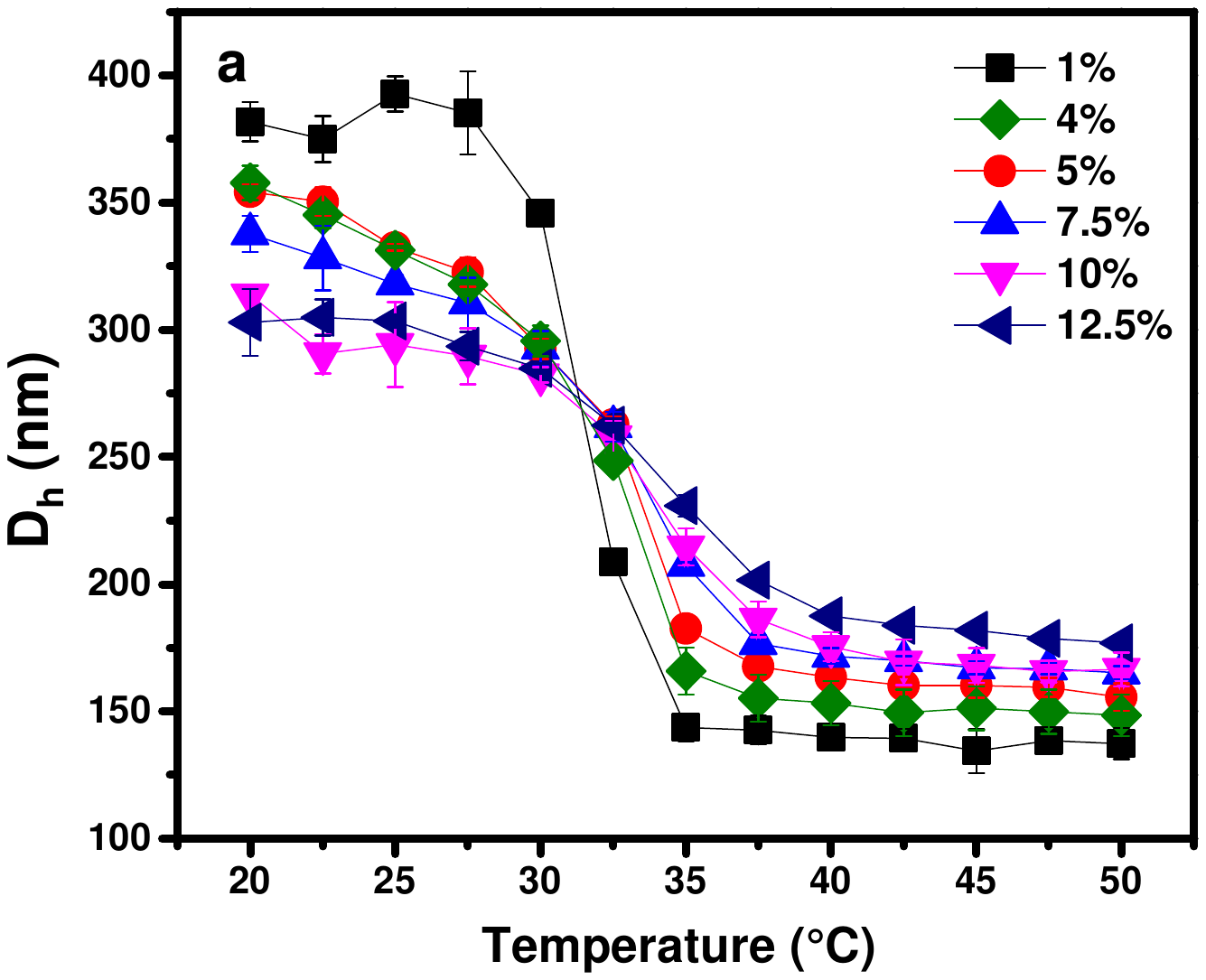}
   \includegraphics[height=6cm]{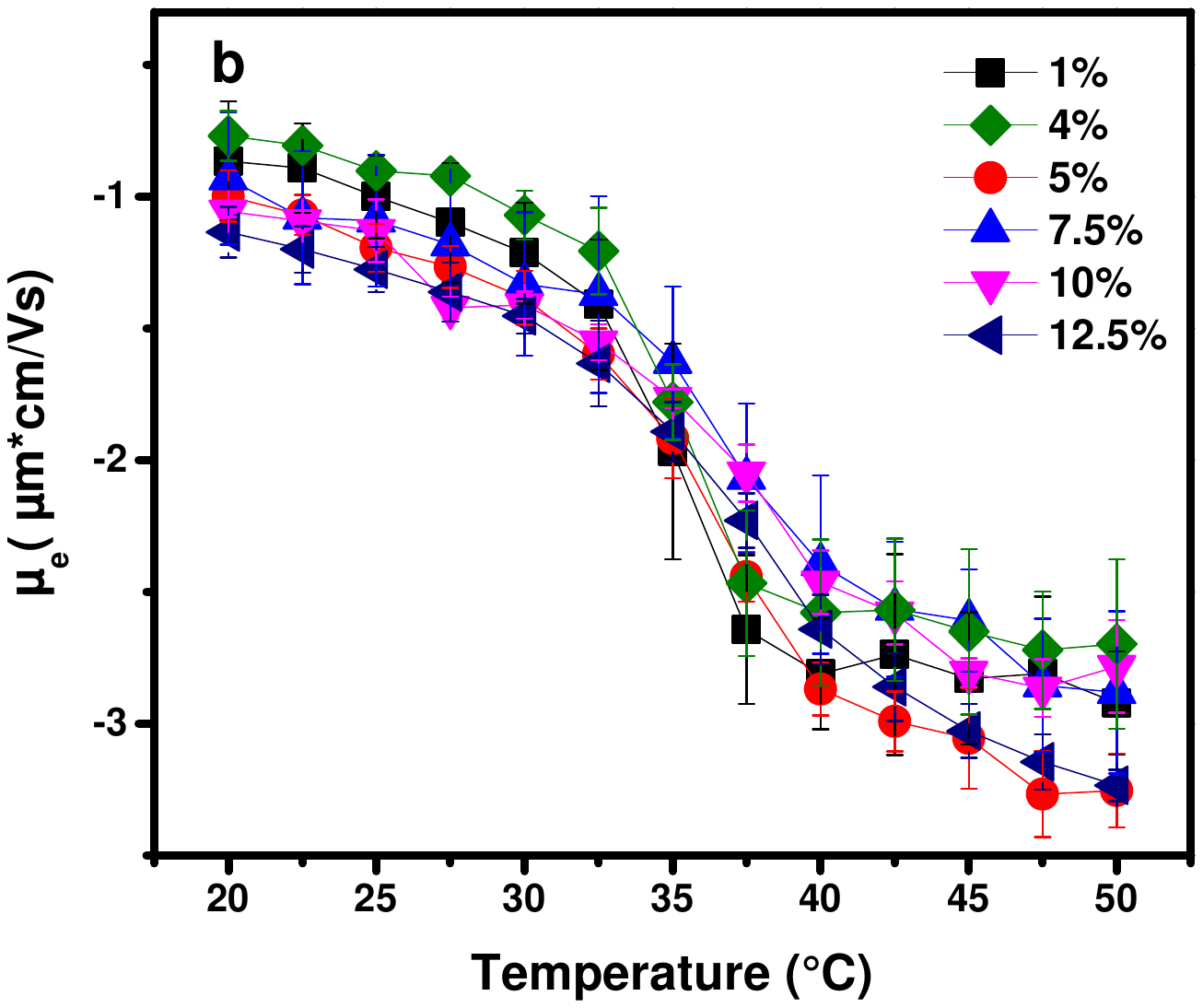}
  \caption{Comparison of hydrodynamic diameter (a) and electrophoretic mobility (b) of pNIPAM microgels with different crosslinking density as a function of temperature}
  \label{mobility}
\end{figure}

 Fig. \ref{mobility}  depicts the temperature-dependent behavior of hydrodynamic diameter (a)  and electrophoretic mobility (b), respectively, for pNIPAM microgels with varying crosslinker concentrations. A clear correlation emerges between crosslinking density and the magnitude of the volume phase transition. Microgels with lower crosslinker content exhibit more pronounced size changes upon heating, as evidenced by the steeper slope in the hydrodynamic diameter profiles. Conversely, microgels with higher crosslinker concentrations demonstrate a less pronounced thermoresponsive behavior, characterized by a shallower slope. Interestingly, the electrophoretic mobility profiles in Fig. \ref{mobility}(b) reveal a different trend. The absolute magnitude of mobility and transisiton slopes are largely independent of crosslinking density.

\begin{figure}[h!]
\centering
  \includegraphics[height=6cm]{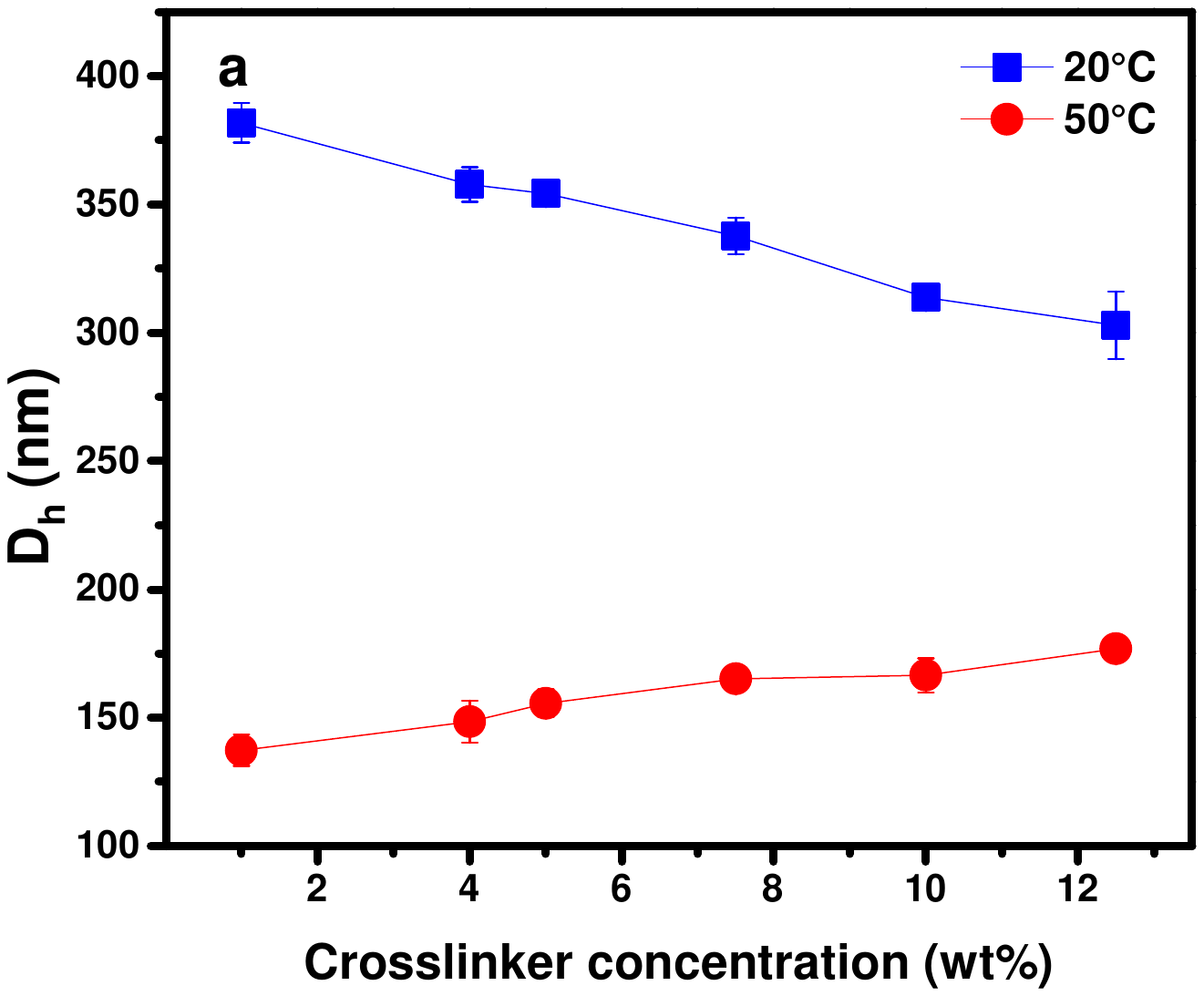 }
   \includegraphics[height=6cm]{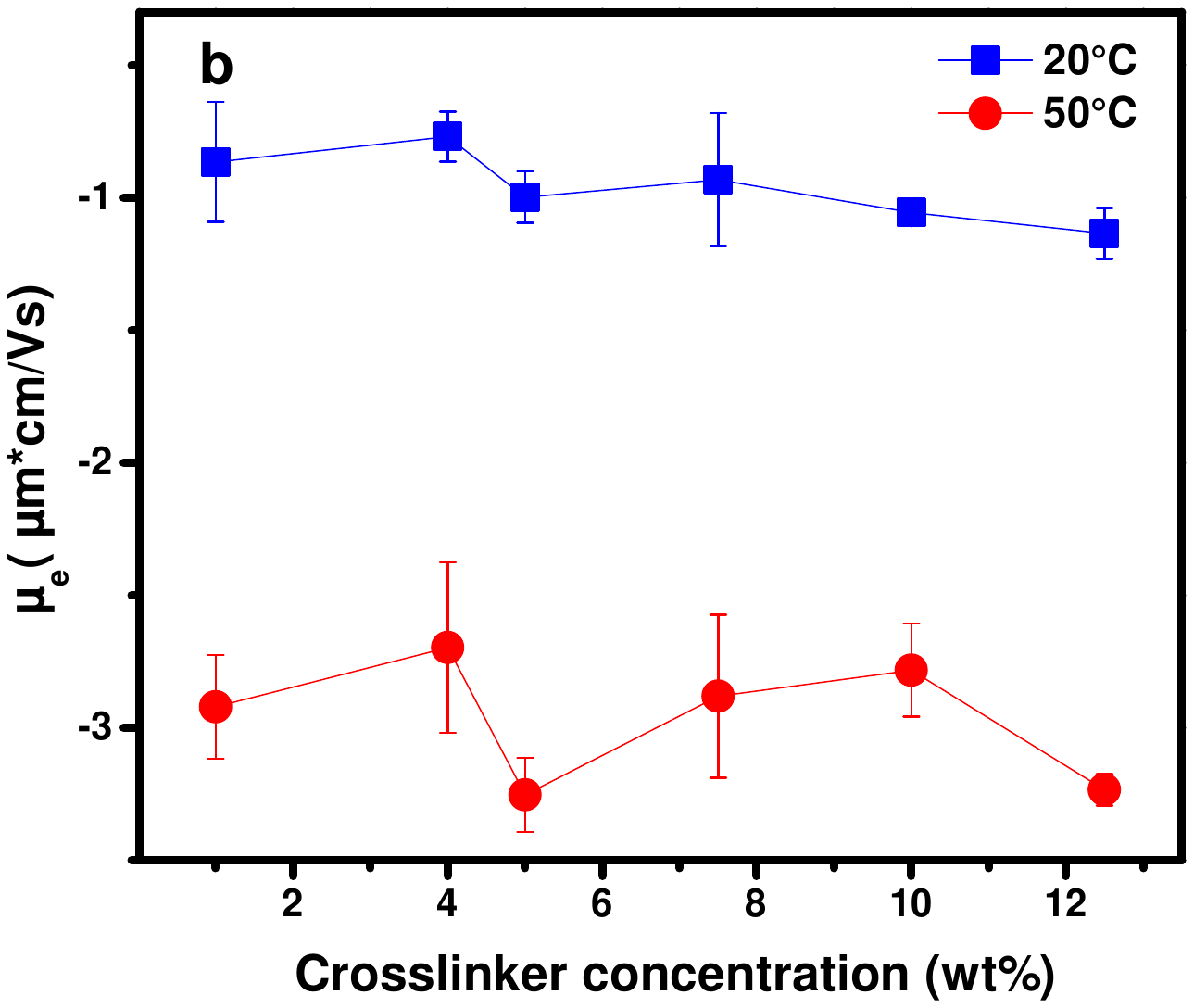}
  \caption{Hydrodynamic diameter (a) and electrophoretic mobility (b)of pNIPAM microgels with different crosslinking densities (1$wt\%$, 4$wt\%$, 5$wt\%$, 7.5$wt\%$, 10$wt\%$, and 12.5$wt\%$) as a function of crosslinker concentration at 20$^{\circ}C$ and 50$^{\circ}C$. Error bars represent standard deviation}
  \label{2050}
\end{figure}

To further explore the effect of crosslinking density at specific temperatures, Fig. \ref{2050} presents the hydrodynamic diameter (a) and electrophoretic mobility (b) as a function of crosslinker concentration at 20$^{\circ}C$ (swollen state) and 50$^{\circ}C$ (collapsed state). As observed in Fig. \ref{2050} (a), the size difference between the swollen and collapsed states diminishes with increasing crosslinker concentration,  indicating a less collapsible network structure at higher crosslinking density. Interestingly, Fig. \ref{2050} (b) reveals that the absolute values of electrophoretic mobility are largely independent of crosslinker concentration at both 20$^{\circ}C$ and 50$^{\circ}C$

\subsection{Estimation of transition temperatures}
\begin{figure}[h!]
\centering
  \includegraphics[height=6cm]{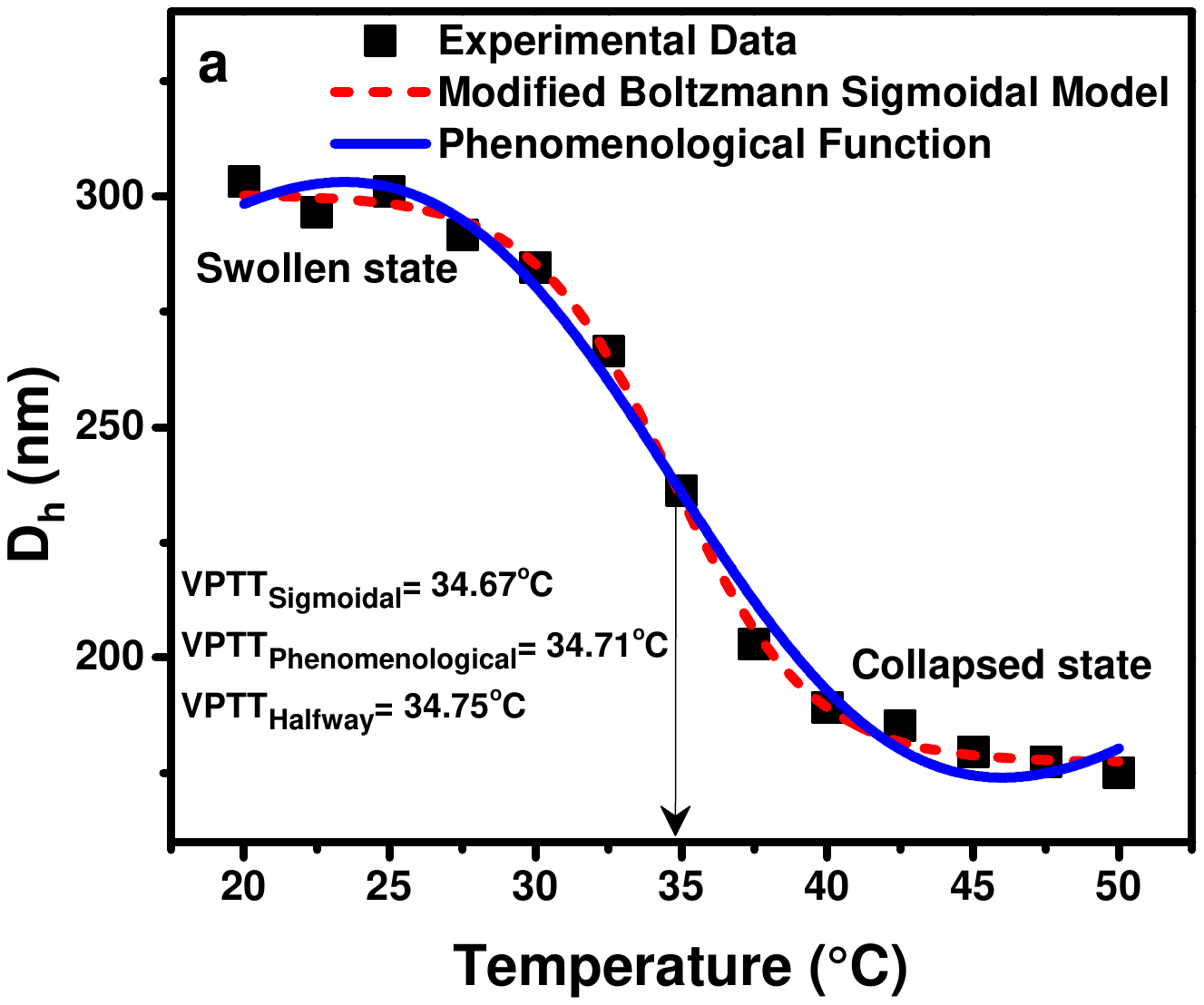}
   \includegraphics[height=6cm]{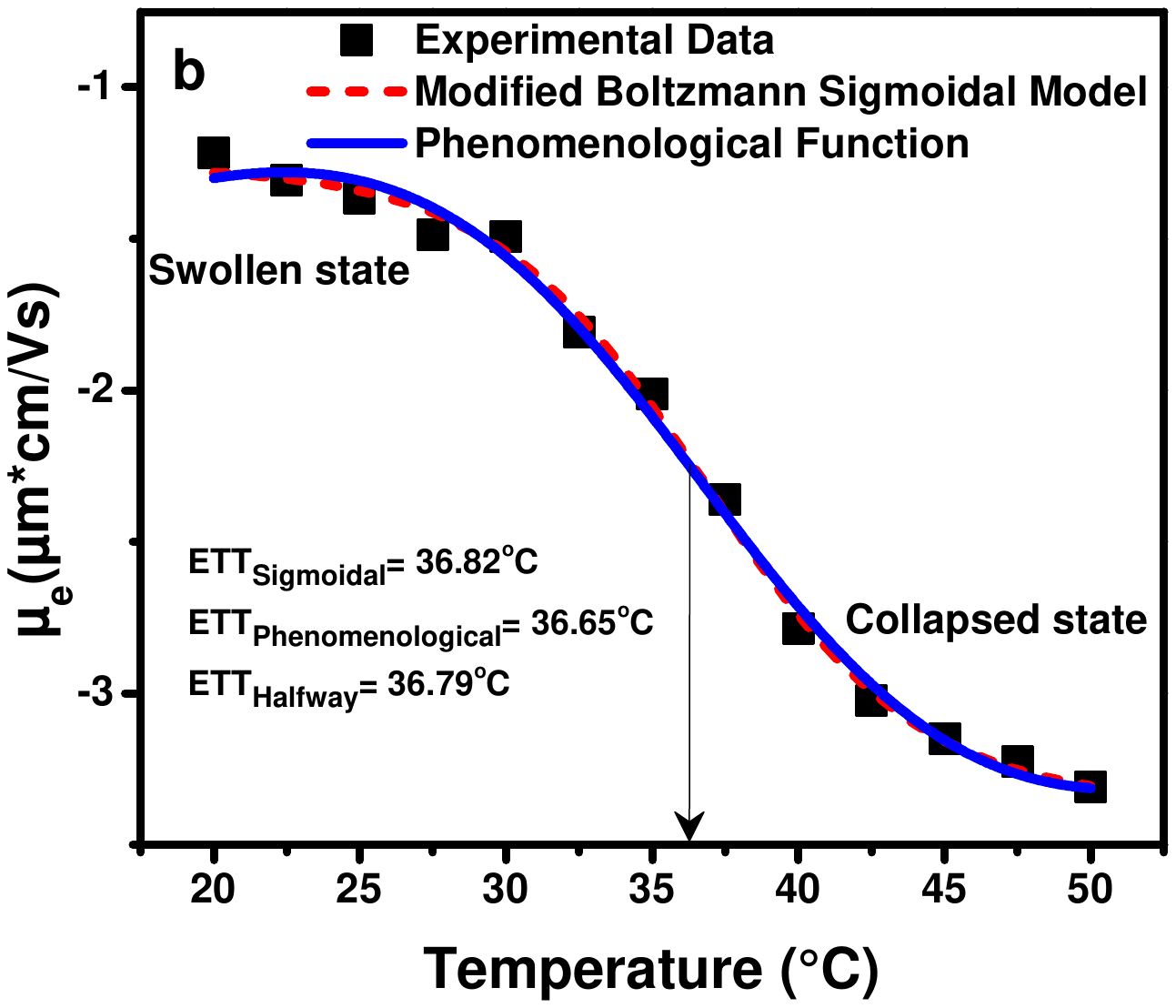}
\caption {Hydrodynamic diameter (a) and electrophoretic mobility (b) of pNIPAM microgels with a crosslinking density of 12.5$wt\%$ are plotted as a function of temperature. The black squares represent the experimental data. The superimposed lines depict the fitting results using two models: the phenomenological function (solid blue line) and the modified boltzmann sigmoidal model (red dash line). Vertical arrows indicate the VPTT in (a) and ETT in (b), determined by three independent methods which are indistinguishable}
  \label{fitting}
\end{figure}

VPTT and ETT for each batch of microgel were determined using three independent methods: phenomenological function, modified boltzmann sigmoidal model, and half-way method. To illustrate the fitting process, a representative temperature sweep experiment for the 12.5$wt\%$ crosslinker concentration microgel batch is shown in Fig. \ref{fitting}. This figure depicts both the hydrodynamic diameter Fig. \ref{fitting} (a) and electrophoretic mobility Fig. \ref{fitting} (b) data plotted against temperature. The data for both parameters were then fitted using the phenomenological function and the modified boltzmann sigmoidal model. Similar figures for other crosslinker concentrations are shown in the supplementary information \cite{si}. As Fig. \ref{fitting} demonstrates, these models achieve good agreement between the experimental data and the fitted curves, highlighting the reliability of these methods in extracting the VPTT and ETT values, while we aimed for reduced R-squared values to be greater than 0.95, it's important to acknowledge that this metric alone might not be sufficient for a comprehensive evaluation of the fitting quality. In some cases, even models with slightly lower R-squared values can be reliable if they capture the essential trends and physical meaning of the data.

\subsection{Effect of crosslinker concentration on transition temperature}
\begin{table*}[h!]
    \centering
    \resizebox{\textwidth}{!}{ 
    \begin{tabular}{|c|c|c|c|c|c|c|c|c|c|c|c|c|c|}
    \hline
    S.No & Crosslinker concentration & \multicolumn{2}{c|}{Phenomenological function} & \multicolumn{2}{c|}{Modified boltzmann sigmoidal model} & \multicolumn{2}{c|}{Halfway method} \\
    \hline
    & ($wt\%$) & ETT ($^{\circ}C$) & VPTT ($^{\circ}C$) & ETT ($^{\circ}C$) & VPTT ($^{\circ}C$) & ETT ($^{\circ}C$) & VPTT ($^{\circ}C$) \\
    \hline
    1 & 1 & 34.45 $\pm$ 0.20 & 31.75 $\pm$ 0.20 & 34.28 $\pm$ 0.54 & 31.70 $\pm$ 0.26 & 34.32 $\pm$ 0.34 & 31.40 $\pm$ 0.16 \\
    \hline
    2 & 4 & 34.88 $\pm$ 0.11 & 32.14 $\pm$ 0.15 & 34.61 $\pm$ 0.34 & 31.58 $\pm$ 0.21 & 34.81 $\pm$ 0.28 & 32.31 $\pm$ 0.25 \\
    \hline
    3 & 5 & 36.46 $\pm$ 0.39 & 32.21 $\pm$ 0.23 & 35.89 $\pm$ 0.49 & 32.02 $\pm$ 0.30 & 36.02 $\pm$ 0.09 & 32.54 $\pm$ 0.26 \\
    \hline
    4 & 7.5 & 37.44 $\pm$ 0.31 & 32.91 $\pm$ 0.01 & 36.43 $\pm$ 0.57 & 32.59 $\pm$ 0.15 & 36.53 $\pm$ 0.25 & 32.96 $\pm$ 0.10 \\
    \hline
    5 & 10 & 36.66 $\pm$ 0.20 & 33.85 $\pm$ 0.11 & 36.30 $\pm$ 0.17 & 34.04 $\pm$ 0.29 & 36.33 $\pm$ 0.03 & 33.56 $\pm$ 0.11 \\
    \hline
    6 & 12.5 & 36.58 $\pm$ 0.37 & 34.34 $\pm$ 0.29 & 37.29 $\pm$ 0.37 & 34.29 $\pm$ 0.29 & 37.04 $\pm$ 0.20 & 34.61 $\pm$ 0.10 \\
    \hline
      \end{tabular}}
    \caption{VPTT and ETT of pNIPAM microgels: This table summarizes the VPTT and ETT values for pNIPAM microgels prepared with different crosslinker concentrations (1$wt\%$, 4$wt\%$, 5$wt\%$, 7.5$wt\%$, 10$wt\%$, and 12.5$wt\%$). The values are reported for the three independent methods used to estimate the transition temperatures: the phenomenological function, the modified boltzmann sigmoidal model, and the half-way method.}
    \label{transistiontemp}
\end{table*}

\begin{figure}[h!]
 \centering
 \includegraphics[height=6cm]{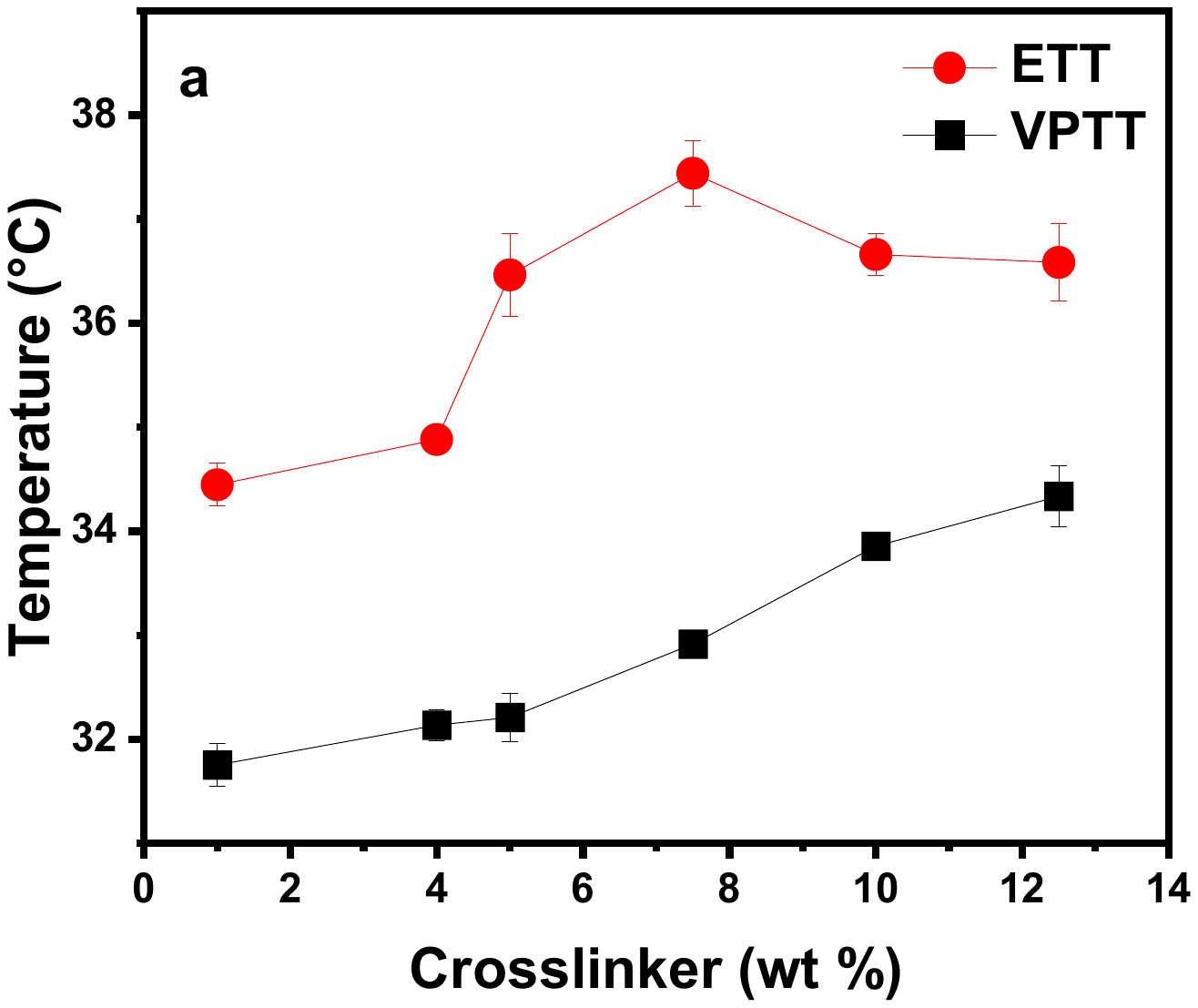}
 \includegraphics[height=6cm]{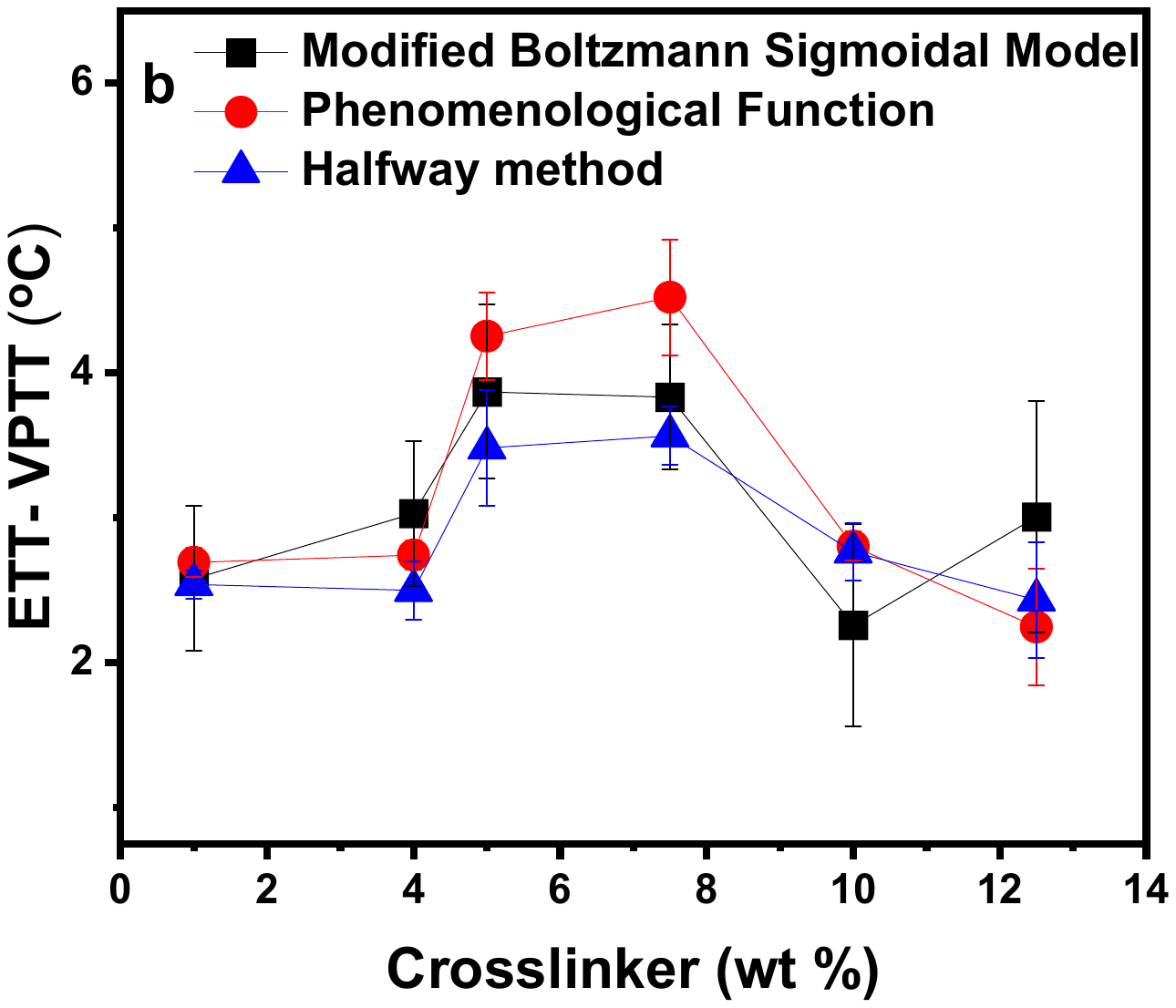}
 \caption{ Influence of crosslinker concentration on transition temperatures: (a) depicts VPTT and ETT of pNIPAM microgels (phenomenological function used to estimate them) plotted as a function of crosslinker concentration. As observed, ETT remains consistently higher than VPTT across all crosslinker concentrations  and (b) illustrates the difference between ETT and VPTT for the pNIPAM microgels at varying crosslinker concentrations. The observed trend is non-monotonic, suggesting a complex interplay between crosslinking density and transition temperatures.}
 \label{vpttettdiff}
\end{figure}

The influence of crosslinker concentration on the thermoresponsive behavior of the microgels was investigated through the determination of VPTT and ETT using three independent methods. The extracted values, summarized in Table \ref{transistiontemp}, reveal a consistent trend where ETT is larger than VPTT across all crosslinker concentrations. Additionally, VPTT exhibits a monotonic increase with increasing crosslinker density and a non-monotonic behavior is evident in the ETT values, with an initial increase followed by a decrease as crosslinker concentration increases. This suggests a complex interplay between crosslinking density, surface charge, and the microgel's response to temperature change. To visualize these trends, Fig. \ref{vpttettdiff} (a) presents the variation of VPTT and ETT with crosslinker concentration, determined using the phenomenological function. Similar trends were observed using the modified boltzmann sigmoidal and halfway methods, presented in the supplementary information \cite{si}.

Fig. \ref{vpttettdiff} (b) depicts that this difference between transition temperatures exhibits a non-monotonic trend with increasing crosslinker concentration. The intermediate crosslinked microgels show the highest difference in transition temperature. In this study, the difference between the VPTT and ETT ranged from 1$^{\circ}C$ to 4.5$^{\circ}C$. This value falls below the range reported by Daly and Saunders\cite{daly2000temperature}, who employed a surfactant-free emulsion polymerization (SFEP) method for microgel synthesis. In contrast, our study utilized a conventional FRP approach. In addition, Daly and Saunders \cite{daly2000temperature} used different concentrations of KCl during their size and electrophoretic mobility analysis. While the absence of salts in our measurements could be one contributing factor, it's important to consider these as potential reasons for this discrepancy in magnitude.

Recalling the earlier discussion, the core of the microgel collapses initially as the temperature approaches VPTT. This phenomenon occurs because the core, being less influenced by electrostatic repulsion, primarily experiences collapse due to thermal-induced alterations in polymer-solvent interactions. As the temperature continues to increase, the corona undergoes collapse at ETT. The intriguing non-monotonic trend observed in the plot of (ETT-VPTT) versus crosslinker concentration Fig. \ref{vpttettdiff} b suggests a complex interplay between charges, crosslinking density and core-corona structure.

Hypothesis:

A two-step collapse mechanism is proposed for pNIPAM microgels, influenced by both crosslinking density and electrostatic interactions. Initially, the microgel core, characterized by a denser polymer network, undergoes a primary collapse driven by thermal-induced alterations in polymer-solvent interactions. Subsequently, the more loosely packed corona, enriched with charged initiator groups, undergoes a secondary collapse influenced by both electrostatic repulsion and polymer-solvent interactions. The interplay between these factors, modulated by crosslinking density, determines the overall deswelling behavior and the magnitude of the VPTT-ETT difference. 

At hypothetically lowest crosslinking density, the core may be very small and the microgel in whole loosely cross-linked. This condition could result in a single-step collapse, where both the core and corona regions deswell simultaneously, leading to a minimal difference between VPTT and ETT. At intermediate cross-Linker Concentration the core is likely to be more prominent and stark difference exists with the loosely cross linked corona. This scenario would facilitate a distinct two-step collapse, characterized by a significant difference between VPTT and ETT, with the core transitioning first, followed by the corona at a higher temperature. At the largest crosslinking density, the entire microgel structure is expected to become excessively rigid, potentially hindering the corona's ability to fully collapse. This could lead to a less pronounced two-step collapse, with both regions collapsing more concurrently, resulting in a smaller difference between VPTT and ETT again and this could be the reason for the non monotonic trend observed in the Fig. \ref{vpttettdiff} (b).

\section{Conclusion}

This study investigated the thermoresponsive behavior of pNIPAM microgels with varying crosslinking densities. We systematically prepared microgels with different 'softness' by adjusting the crosslinker concentration while maintaining all other ingredients constant, including the initiator concentration, which determines the microgel's surface charge density. To determine the size and surface charge response of the microgels as a function of temperature, we performed a combination of electrophoretic mobility (ELS) and dynamic light scattering (DLS) measurements. The volume phase transition temperature (VPTT) and the electrokinetic transition temperature (ETT) for each microgel batch have been determined using three separate techniques: the Phenomenological function, the Modified Boltzmann sigmoidal model, and the half-way method. Our main conclusion is that, crosslinking density primarily affects the magnitude of the volume phase transition, transition temperatures VPTT and ETT,  rather than the absolute values of electrophoretic mobility. In other words, the softness of the microgel influences how it transitions from a swollen to a collapsed state but not the inherent mobility of the charged groups at a specific temperature. Furthermore, the difference between ETT and VPTT shows a non-monotonic trend as crosslinking density increases. 
The absence of salts in our measurements compared to previous studies might have contributed to the observed differences in transition temperatures. These findings provide valuable insights into the interplay between crosslinking density and the thermoresponsive behavior of pNIPAM microgels.  
To gain a deeper understanding of the underlying mechanisms governing the observed trends, future studies should employ advanced characterization techniques such as small-angle neutron scattering (SANS) to investigate the internal structure, albeit the size of core and corona, of an individual microgel as a function of crosslinking density. Additionally, theoretical modeling and simulations incorporating the effects of crosslinking density, electrostatic interactions, and solvent quality could provide valuable insights into the factors driving the observed thermoresponsive behavior.
\section*{Author contributions}
Syamjith KS: Conceptualization and implementation of methodology, data analysis and developing hypothesis, writing – original draft, review, \& editing.\\
Shubhasmita Rout: Synthesis of pNIPAM microgels with varying crosslinking densities.\\
Alan R Jacob: Project administration, supervision, funding acquisition, writing – review, \& editing.

\section*{Conflicts of interest}
There are no conflicts to declare.

\section*{Acknowledgements}
The authors gratefully acknowledge the invaluable support of the Soft Matter Group (SMG) lab members throughout this project. Their assistance and collaborative spirit significantly contributed to the research. The authors acknowledge the financial support provided by the Science and Engineering Research Board (SERB) for the project (SRG/2021/000779), and ARJ specifically acknowledges the seed grant provided by IIT Hyderabad. The funding enabled us to acquire the necessary materials and equipment, allowing us to conduct this research. We are particularly grateful to Dr. Domenico Truzzolillo for his insightful discussions and critique on this manuscript. The authors also extend their appreciation to the team members of Polymer Engineering and Colloid Science (PECS) at IIT Madras for facilitating measurements and engaging in stimulating discussions. ARJ dedicates this manuscript to Prof. Stefan U. Egelhaaf an amazing scientist more importantly, a wonderful human being. 



\balance


\bibliography{rsc} 
\bibliographystyle{rsc} 

\end{document}